\documentclass[10pt]{iopart}
\usepackage{bm}
\usepackage{graphicx}
\usepackage[usenames]{color}
\usepackage{microtype}
\usepackage{amsmath}
\usepackage{amssymb}
\usepackage{xspace}
\usepackage{footnote}
\usepackage{relsize}
\usepackage{hhline}

\newcommand{\bb}{\mathbf{b}}

\newcommand{\bc}{\mathbf{c}}

\newcommand{\bv}{\mathbf{v}}
\newcommand{\bV}{\mathbf{V}}
\newcommand{\bS}{\mathbf{S}}

\newcommand{\bK}{\mathbf{K}}
\newcommand{\bPsi}{\mathbf{\Psi}}

\newcommand{\sN}{{N}}
\newcommand{\sM}{{M}}
\newcommand{\stM}{{D}}

\usepackage{physics}
\usepackage[normalem]{ulem}
\usepackage{hyperref}

\begin{document}

\title{Electronic-structure properties from atom-centered predictions of the electron density}

\author{Andrea Grisafi$^1$, Alan M.~Lewis$^2$, Mariana Rossi$^2$ and  Michele Ceriotti$^3$}

\address{$^1$ PASTEUR, Département de Chimie, \'Ecole Normale Supérieure, 75005 Paris, France}
\address{$^2$ Max Planck Institute for the Structure and Dynamics of Matter, Luruper Chaussee 149, 22761 Hamburg, Germany}
\address{$^3$ Laboratory of Computational Science and Modeling, IMX, \'Ecole Polytechnique F\'ed\'erale de Lausanne, 1015 Lausanne, Switzerland}

\ead{andrea.grisafi@ens.psl.eu, alan.lewis@mpsd.mpg.de}

\begin{abstract}
The electron density of a molecule or material has recently received major attention as a target quantity of machine-learning models. A natural choice to construct a model that yields transferable and linear-scaling predictions is to represent the scalar field using a multi-centered atomic basis analogous to that routinely used in density fitting approximations. However, the non-orthogonality of the basis poses challenges for the learning exercise, as it requires accounting for all the atomic density components at once.  We devise a gradient-based approach to directly minimize the loss function of the regression problem in an optimized and highly sparse feature space. In so doing, we overcome the limitations associated with adopting an atom-centered model to learn the electron density over arbitrarily complex datasets, obtaining extremely accurate predictions. The enhanced framework is tested on 32-molecule periodic cells of liquid water, presenting enough complexity to require an optimal balance between accuracy and computational efficiency. We show that starting from the predicted density a single Kohn-Sham diagonalization step can be performed to access total energy components that carry an error of just 0.1 meV/atom with respect to the reference density functional calculations. Finally, we test our method on the highly heterogeneous QM9 benchmark dataset, showing that a small fraction of the training data is enough to derive ground-state total energies within chemical accuracy.
\end{abstract}

\ioptwocol
\section{Introduction}

In the last few years, many efforts have been devoted to the integration of supervised learning techniques within state-of-the-art electronic-structure methods, aiming at accelerating the calculation of properties beyond ground-state total energies and forces~\cite{harmann+20nc,Westermayr2021,kirkpatrick+21science,Kulik2022}. Different quantities  have been considered as the target of machine learning (ML) models, including molecular multipoles~\cite{bere+15jctc,Staacke2022,veit+20jcp} and  polarizabilities~\cite{raim+19njp}, electronic density of states and bandgaps~\cite{Pilania2017,mahmoud+20prb,Westermayr2021cr}, kinetic density functionals~\cite{snyd+12prl,meye+20jctc,Ryczko2022}, exchange-correlation potentials~\cite{Zhou2019,Schmidt2019,Suzuki2020}, single-particle wavefunctions~\cite{welb+18jctc,schu+19nc,Gastegger2020} and  Hamiltonians~\cite{schu+19nc,west-maur21cs,nigam+21jcp}. Among those, a special role is played by the electron density $\rho$, which, unlike other ingredients of electronic-structure methods, represents a well-defined physical observable that can be experimentally measured~\cite{korits+01cr}. 

From a theoretical point of view, the importance of a learning model that provides inexpensive predictions of the electron density is apparent when used in conjunction with density-functional theory (DFT). Within DFT, $\rho$ provides direct access to not just the ground-state electronic energy, but in principle to all of the ground-state properties of a system. While this sounds an attractive possibility, it is not always obvious whether one can actually benefit from computing derived electronic properties from the predicted densities. In fact, under many circumstances it appears more convenient to target the prediction of the property at hand directly, without going through intermediate electronic-structure ingredients~\cite{gris+19acscs}. Importantly, however, it has been recently shown that using the electron density as a stepping stone to predict the electronic energy of a system can be beneficial when working in highly extrapolative regimes, which are specifically associated with a substantial increase of the system size~\cite{lewis+21jctc}. This behaviour is to be expected, especially when the machine-learning approximation lacks a description of long-range interactions between two distant portions of the system, so that computing the electronic energy as a functional of $\rho$ can effectively recover electrostatic and non-local effects. For these reasons, designing efficient and accurate learning algorithms that specifically target the prediction of the electron density is becoming an increasingly attractive possibility both to provide useful insights on the electronic structure of the system and to complement the predictions of local ML potentials.

The main difficulty associated with the regression of the electron density consists in adapting the learning model to a suitable discretization of the scalar field that can be numerically treated by a computer. In fact, the various density-learning models currently available primarily differ by the nature of this discretization procedure, going from working with a finite number of real-space grid points~\cite{alre+18cst,chan+19npjcm,Jorgensen2020,Jorgensen2021} to expanding the density over a suitable set of basis functions~\cite{broc+17nc,bogo+20nc,gris+19acscs,Rackers2022}. In line with previous works, we will consider here a linear atom-centered expansion of the density analogous to that of density-fitting (DF) approximations routinely used in quantum chemistry codes~\cite{bearends1973,weigend+06pccp,golze+17jctc,ren+12njp}. In particular, we will refer to an extension of the density-learning model reported in Ref.~\cite{lewis+21jctc}, which can be used to treat on the same footing both molecular and periodic condensed-phase systems.  We named this method the \textit{symmetry-adapted learning of three-dimensional electron densities} (SALTED). 

Within SALTED, a kernel-based parametrization of the learning problem is provided through a local representation of the atomic structure that mirrors the rotational symmetry of the atomic density components. The atom-centered nature of the density expansion is a critical aspect to ensure the transferability of the SALTED predictions across atomic environments that share similar local structural patterns. The non-orthogonal character of common DF basis sets, however, implies that particular care must be given to the spatial overlap between any pair of off-centered basis functions, which has the effect of coupling all the atomic density components together. In fact, the problem dimensionality depends not only on the number of atomic environments used to train the model but also on the basis set size associated with each atomic center. This downside of the method hinders the calculation of explicit solutions at an increasing number of training structures, effectively limiting the application of SALTED to relatively small datasets. Furthermore, as the number of training atomic environments can grow very quickly, suitable dimensionality reduction schemes must be adopted in order to limit the size of the problem to be solved. In this respect, converging the prediction accuracy with respect to the number of sparse atomic inputs is known to be a particularly hard numerical problem which requires the implementation of suitable expedients~\cite{foster+09jmlr}. These shortcomings become particularly severe whenever a large number of atomic density components are required to train the model, e.g., when targeting the prediction of the electron density of proteins~\cite{fabr+20chimia} or generic disordered, liquid systems. 

In this paper, we overcome the aforementioned limitations in two ways: i) we make use of a featurization of the atomic structure expressed explicitly in the \textit{reproducing kernel Hilbert space} (RKHS) of symmetry-adapted sparse kernel approximations, thus enabling a systematic convergence of the prediction accuracy with the number of sparse atomic environments,
and ii) we implement an iterative gradient-based minimization of the SALTED loss function, which allows us to bypass the computational bottleneck associated with the inversion of prohibitively large regression matrices. Taking bulk liquid water as an example, we show that both these aspects combined allow us to apply SALTED to arbitrarily complex systems, paving the way for the inexpensive calculation of electronic structure properties and accurate electronic energies.

\section{Theory}

Consider a generic density-fitting approximation of the electron density of a given structure $A$, expressed as a linear expansion over a set of atom-centered basis functions $\phi_{an\lambda\mu}$:
\begin{equation}
\label{eq:rho-non-orthogonal}
\rho_A^\text{DF}(\mathbf{r})  =\sum_{i,an\lambda\mu,\mathbf{U}} c_{an\lambda\mu}(A_i)\, \phi_{an\lambda\mu}(\mathbf{r}- \mathbf{R}_i + \mathbf{U})\, .
\end{equation}
$\mathbf{R}_i$ labels the position of atom $i$, $\mathbf{U}$ is the cell translation vector, assuming the system is periodic, and $c_{an\lambda\mu}(A_i)$ are the non-orthogonal expansion coefficients. The basis $\phi_{an\lambda\mu}$ is defined as the atom-centered product of spherical harmonics $Y_{\lambda}^\mu$ and a set of radial functions $R^{a\lambda}_{n}$ which depend on the possible choices of atomic types $a$ and angular momenta $\lambda$. Similarly to what is reported in previous works~\cite{gris+19acscs,fabr+19cs,lewis+21jctc}, we then provide an atom-centered approximation to the expansion coefficients through the definition of a covariant similarity measure between atomic environments, which takes into account the transformation properties of spherical harmonics under rotations:
\begin{equation}\label{eq:coeffs_approx}
    c_{a n\lambda\mu}(A_i) \approx \sum_{\{M_j^{a n\lambda}\}}\sum_{|\mu'|\leq \lambda} b^{a n\lambda}_{j\mu'} \operatorname{k}^{a n\lambda}_{\mu\mu'}(A_i,M^{a n\lambda}_j) .
\end{equation}
Here, $b^{a n\lambda}_{j\mu'}$ are the regression weights and $\operatorname{k}^{a n\lambda}_{\mu\mu'}(A_i,M^{a n\lambda}_j)$ is a symmetry-adapted kernel function~\cite{gris+18prl} that couples the atomic environment $A_i$ of the training set with a subset of atomic environments $\{ M_j^{a n\lambda} \}$ selected to provide a suitable sparse representation of the spectrum of chemical and structural variations included in the training set. Note that in the definition of the sparse set $\{ M_j^{a n\lambda} \}$ we left the freedom of selecting different atomic environments depending on the atomic function type $a n\lambda$. In contrast to previous implementations, we investigate the possibility of using different kernels not only to reflect the different types of angular symmetries associated with each set of spherical harmonics $Y_{\lambda}^\mu$, but also to reflect the different length scales associated with the various radial functions $R^{a\lambda}_{n}$. Each kernel is constructed from a corresponding $\lambda$-SOAP representation~\cite{gris+18prl} computed at an optimized spatial resolution.  

Inserting Eq.~\eqref{eq:coeffs_approx} into Eq.~\eqref{eq:rho-non-orthogonal}, a ML approximation of the electron density  $\rho^\text{ML}$ can be defined to linearly depend on the vector of regression weights $\bb_{\sM}$, where we use the subscript $\sM$ to indicate that the vector corresponds to the full set of reference environments. 
The vector $\bb_\sM$ can then be obtained from the minimization of a quadratic loss function that measures the integrated error over the $N$ training structures between $\rho^\text{ML}$ and the reference DF definition of Eq.~\eqref{eq:rho-non-orthogonal}:
\begin{equation}\begin{split}\label{eq:loss}
\ell(\bb_\sM) &= \sum^N_{A=1}\int_{u.c.} d\mathbf{r} \left| \rho^\text{ML}_A(\mathbf{r};\bb_\sM) - \rho_A^{\text{DF}}(\mathbf{r}) \right|^2 \\& + \eta\,\bb_\sM^T\mathbf{K}_{\sM \sM}\bb_\sM\, ,
\end{split}\end{equation}
where in the second line we introduced a regularization term modulated by a hyperparameter $\eta$. The matrix $\mathbf{K}_{\sM \sM}$ includes all the symmetry-adapted kernels $\operatorname{k}^{an\lambda }_{\mu\mu'}(M^{an\lambda}_i,M^{an\lambda}_j)$ that couple the sparse set of atomic environments $\{ M_j^{a n\lambda} \}$ and it is thus defined to be block diagonal in the possible basis function types~$\{a n\lambda \}$.

\subsection{RKHS formulation}
According to standard \textit{subset of regressors}~\cite{quin+05jmlr} (SoR) approaches, selecting the number of sparse atomic environments $M^{an\lambda}$ as a proper subset of the total number of training environments implies approximating the full kernel matrix of the SALTED problem as follows:
\begin{equation}\label{eq:sor}
    \bK_{\sN\sN} \approx \bK_{\sN\sM}\bK_{\sM\sM}^{-1}\bK_{\sN\sM}^T
\end{equation}
where we used the subscript $\sN$ to indicate that the dimension of $\bK_{\sN\sN}$ is given by the sum of the number of DF basis functions associated with each atom of the training set. 
Most of the time, the matrix $\bK_{\sM\sM}$ is found to be ill-conditioned, yielding severe numerical instabilities when solving the regression problem. Commonly adopted solutions to circumvent this issue consist in introducing an \textit{ad hoc} jitter term that provides a lower bound to the eigenvalues of the matrix to be inverted~\cite{lewis+21jctc}. While effectively fixing the numerical instabilities, this stratagem carries the drawback of hindering the systematic convergence of the SoR approximation when taking the limit of $M^{an\lambda}$ towards the total number of training environments. %
To solve this problem, we first consider a truncation %
of the eigen-decomposition of the $\bK_{MM}$ matrix.
Given that $\bK_{MM}$ is block-diagonal in the basis function type, for each triplet $(a n\lambda)$ we can write
\begin{equation}
   \bK^{a n\lambda}_{\sN \sN} \approx \bK^{a n\lambda}_{\sN \sM} \left(\sum_d^{D^{a n\lambda}}\frac{\bv^d_\sM(\bv^d_\sM)^{T}}{\lambda_d} \right)(\bK^{a n\lambda}_{\sN \sM})^T\, ,
\end{equation}
where $\lambda_d$ are the non-negligible eigenvalues of $\bK^{a n\lambda}_{\sM\sM}$ and $\bv^d_\sM$ the corresponding eigenvectors, so that $D^{a n\lambda}$ indicates the final truncated dimension. Note that each outer product $\bv_{\sM}^d(\bv_{\sM}^d)^{T}$ conserves the transformation properties of $\bK^{a n\lambda}_{\sM\sM}$, so that, individually, each eigenvector $\bv^d_{\sM}$ has the same rotational symmetry as the corresponding basis functions $(a\lambda n)$.
From the equation above, we can then define the \textit{reproducing kernel Hilbert space} (RKHS) of the sparse kernel approximation as the space spanned by the following feature vectors:
\begin{equation}\label{eq:psi-def}
    \bPsi^{a n\lambda}_{\sN \stM} \equiv \bK^{a n\lambda}_{\sN \sM} \bV^{a n\lambda}_{\sM \stM} (\mathbf{\Lambda}^{a n\lambda}_{\stM\stM})^{-\frac{1}{2}}\, ,
\end{equation}
which, consistently with Mercer's theorem~\cite{rasm05book}, gives
\begin{equation}
    \bK^{a n\lambda}_{\sN \sN} \approx \bK^{a n\lambda}_{\sN \sM}\,(\bK^{a n\lambda}_{\sM \sM})^{-1}\,(\bK^{a n\lambda}_{\sN\sM})^T  \approx \bPsi^{a n\lambda}_{\sN\stM}(\bPsi^{a n\lambda}_{\sN\stM})^T\, .
\end{equation}
Finally, we can reformulate the density-learning problem as a linear regression task parametrized according to the feature vectors $\bPsi^{a n\lambda}_{\sN\stM}$. To do so, we can first invert Eq.~\eqref{eq:psi-def} to approximate the $\bK^{a n\lambda}_{\sN\sM}$ matrices~as
\begin{equation}
   \bK^{a n\lambda}_{\sN \sM}  \approx \bPsi^{a n\lambda}_{\sN \stM} (\mathbf{\Lambda}^{a n\lambda}_{\stM\stM})^{\frac{1}{2}} (\bV^{a n\lambda}_{\sM\stM})^T \, ,
\end{equation}
where we made use of the unitary nature of the matrix of eigenvectors, $\bV^{a n\lambda}_{\sM\stM}(\bV^{a n\lambda}_{\sM\stM})^T=\boldsymbol{1}^{a n\lambda}_{\sM\sM}$. Then, a sparse RKHS approximation of the density expansion coefficients can finally be obtained by rewriting Eq.~\eqref{eq:coeffs_approx} as follows
\begin{equation}\label{eq:rkhs-coeffs_approx}
    c_{an\lambda\mu}(A_i)  \approx \sum_{d}^{D^{an\lambda}}  \tilde{b}^{a n\lambda}_d \psi_d^{an\lambda}(A_i;an\lambda\mu)\, ,
\end{equation}
where, for each basis function type $(a n\lambda)$, we defined a new vector of regression weights as
\begin{equation}
    \tilde{\bb}^{a n\lambda}_{\stM} \equiv (\mathbf{\Lambda}^{a n\lambda}_{\stM\stM})^{\frac{1}{2}} (\bV^{a n\lambda}_{\sM\stM})^T \bb^{a n\lambda}_{\sM}\, .
\end{equation}
The so derived RKHS reformulation of SALTED allows us to effectively make use of a sparse symmetry-adapted kernel approximation of the density expansion coefficients that avoids the numerical instabilities associated with normal equations solutions~\cite{foster+09jmlr}. 

\subsection{Iterative solution}

\begin{figure}[t!]
    \centering
    \includegraphics[width=8cm]{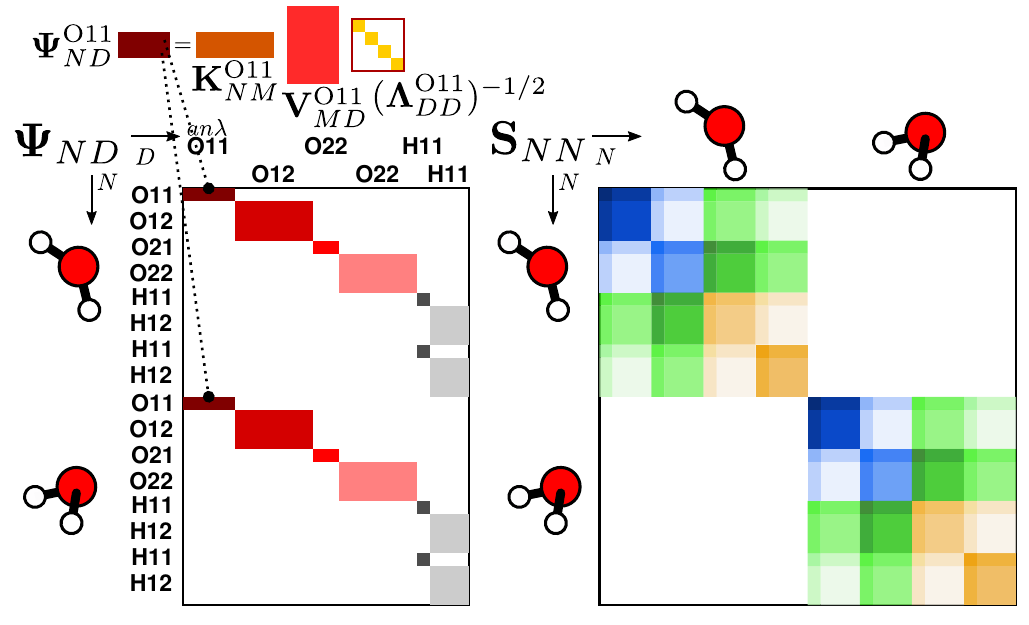}
    \caption{Schematic representation of the sparsity pattern of the RKHS feature matrix $\boldsymbol{\Psi}_{ND}$ (\textit{left}) and basis-set overlap matrix $\boldsymbol{S}_{NN}$ (\textit{right}) entering the iterative minimization of the electron-density loss function.
    $\boldsymbol{\Psi}_{ND}$ can be built from separate blocks, each associated with a separate type of atomic basis functions. 
    } %
    \label{fig:grad-blocks}
\end{figure}

To practically solve the SALTED-RKHS problem, it is convenient to define the global problem dimensionality $D$ as the sum of all the truncated dimensions $D^{an \lambda}$ associated with each basis function type.  In so doing, a single vector of regression weights $\tilde{\bb}_{D}$ constructed by stacking together each of the individual $\tilde{\bb}_{D}^{an\lambda}$ can be used to compactly write the SALTED-RKHS approximation of the density coefficients  as
\begin{equation}
c_{an\lambda\mu}(A_i) \approx \tilde{\bb}^T_{D}\Psi_{D}(A_i;an\lambda\mu)\, ,
\end{equation}
where the global feature vector $\Psi_{D}(A_i;an\lambda\mu)$ is defined to be diagonal in the various basis function types.
When inserted into Eq.~\eqref{eq:rho-non-orthogonal}, this allows us to rewrite the loss function of Eq.~\eqref{eq:loss} as follows:
\begin{equation}\label{eq:loss-2}\begin{split}
     \ell(\tilde{\bb}_{D})=&(\boldsymbol{\Psi}_{\sN D}\tilde{\bb}_D-\bc^\text{DF}_\sN)^T\, \bS_{\sN\sN}\, (\boldsymbol{\Psi}_{\sN D}\tilde{\bb}_D-\bc^\text{DF}_\sN)\\&+\eta\tilde{\bb}_D^T\tilde{\bb}_D \, .
\end{split}\end{equation}
with $\bc^\text{DF}_N$ the vector of training DF expansion coefficients.
The matrix $\bS_{NN}$, which is block diagonal in the training set, contains the spatial overlaps between the basis functions of each structure, $\braket{\phi}{\phi'}$, and is responsible for coupling all the basis function types and therefore all the atomic density components together. Importantly, the sparsity pattern of $\boldsymbol{\Psi}_{\sN D}$ differs from that of $\bS_{NN}$, implying that the product $\boldsymbol{\Psi}_{ND}^T\boldsymbol{S}_{NN}\boldsymbol{\Psi}_{ND}$ is, generally, a full matrix. A schematic representation of this is reported in Fig.~\ref{fig:grad-blocks}.

An explicit minimization of the loss function of Eq.~\eqref{eq:loss-2} would be hindered by the quickly growing dimensionality of the problem when considering an increasing number of sparse atomic environments, which would make the inversion of the resulting regression matrix prohibitively expensive when considering a large number of sparse environments and/or a large basis set size. For this reason, we implement a basic conjugate gradient (CG) algorithm to iteratively minimize the loss function of Eq.~\eqref{eq:loss-2} directly. We rely on the analytical definition of the gradient of Eq.~\eqref{eq:loss-2} with respect to $\tilde{\bb}_D$, i.e.,
\begin{equation}\label{eq:gradient}
     \frac{\partial\ell(\tilde{\bb}_D)}{\partial \tilde{\bb}_D}=\Psi_{\sN D}^T\, \bS_{\sN\sN}\, (\Psi_{\sN D}\tilde{\bb}_D-\bc^\text{DF}_\sN)+\eta\tilde{\bb}_D \, ,
\end{equation}
which is updated at each CG step from the knowledge of the Hessian matrix:
 \begin{equation}\label{eq:hessian}
     \mathbf{H}_{DD}=\bPsi_{\sN D}^T\bS_{\sN\sN}\bPsi_{\sN D}+\eta\mathbf{1}_{DD}\, .
 \end{equation}
Note that as the dimension $D$ can easily be $>10^5$ when treating complex datasets, precomputing $\mathbf{H}_{DD}$ would imply running into similar issues to those associated with solving the regression problem explicitly. For this reason, we estimate the gradient and curvature on the minimization surface at each CG step  by directly computing the projection of Eq.~\eqref{eq:hessian} on the CG search direction. %
In this work, an efficient implementation of the latter quantity is obtained by exploiting the sparsity of the feature vector $\boldsymbol{\Psi}_{ND}$ and by parallelizing the sum over the training set $N$ with preloaded overlap matrices.

\subsection{Orthogonal approach}

As a further justification of the need to tackle the non-orthogonal regression problem, we consider an alternative path that could in principle be undertaken to bypass the computational difficulties discussed above. A suitable orthogonal transformation could be introduced in order to recast the SALTED problem into as many independent regression tasks as the number of the different basis function types $\{a n\lambda\}$. In particular, one could formally rewrite the DF approximation of the electron density in terms of set of orthogonal basis functions, defined as follows
\begin{equation}
 \hat{\phi}_{k}(\mathbf{r}) = \sum_{k'}\left[\bS^{-\frac{1}{2}}\right]^{k'}_{k}\phi_{k'}(\mathbf{r})\, ,
\end{equation}
with $\bS^{-\frac{1}{2}}$ the orthogonalization matrix. This transformation is known as L\"owdin orthogonalization~\cite{lowd1950jcp} and it is typically used in the context of single-particle theories of the electronic wavefunction to work in the space of \textit{atomic natural orbitals}~\cite{almof1991aqc}, as well in the localization procedure of maximally-localized Wannier functions~\cite{marz-vand97prb,Marzari2003psik}.
Under this transformation, the orthogonal expansion coefficients can be obtained from the non-orthogonal DF coefficients as
\begin{equation}\label{eq:ortho-coefs}
 \hat{\bc} = \bS^{\frac{1}{2}}\bc \, .
\end{equation}
Unlike hierarchical orthogonalization schemes, such as Gram-Schmidt, the  L\"owdin orthogonalization preserves both the locality of the basis functions about the atomic positions and the symmetry properties of spherical harmonics.
As such, for each of the distinct triplets ($a n \lambda$) that labels the new set of orthogonal basis functions,  a ML approximation of Eq.~\eqref{eq:ortho-coefs} can be defined to follow the same sparse RKHS formulation already reported in Eq.~\eqref{eq:rkhs-coeffs_approx}.

As the basis functions are now orthogonal to each other, $\bS_{NN}=\boldsymbol{1}_{NN}$ in Eq.~\eqref{eq:loss-2} and the  regression problem decouples, leaving as many independent solutions as the number of orthogonal basis function types $\{a n \lambda  \}$. This means that one can afford to compute the regression weights for each $(a n \lambda )$ explicitly, i.e., 
\begin{equation}
\tilde{\bb}_{D}^{a\lambda n} = ((\mathbf{\Psi}_{{N}{D}}^{a\lambda n})^T \mathbf{\Psi}_{{N}{D}}^{a\lambda n} + \eta\mathbf{1}_{{D}{D}})^{-1}\, (\mathbf{\Psi}_{{N}{D}}^{a\lambda n})^T\, \hat{\mathbf{c}}_{N}^{a\lambda n}\, . 
\label{eq:weights}
\end{equation}
Note that because the learning target still remains the entire density field, the value of the regularization parameter $\eta$ is defined to be the same for each independent solution. 

In spite of the clear computational advantages, the presented orthogonal approach has two main downsides. First, since the basis set of each structure now depends on the specific atomic configuration that determines the L\"owdin orthogonal transformation, the resulting regression task is effectively required to learn this transformation in turn, making the method more data hungry. Moreover, reconstructing the density field from the predicted density coefficients necessarily requires to undo the orthogonal transformation as $\bc = \bS^{-\frac{1}{2}}\hat{\bc}$, adding a considerable computational burden at the prediction level. In fact, especially when targeting the prediction of systems with a large number of atoms, the calculation of $\bS^{-\frac{1}{2}}$ can be particularly expensive, undermining the rationale for applying a ML model in the first place.

\subsection{Accuracy metrics}
As already discussed in Ref.~\cite{lewis+21jctc}, the mean spherical average of the density over the dataset, $\bar{\rho}_0(\mathbf{r})$, is used as a baseline field for the learning of $\rho(\mathbf{r})$. This implies that very much like when doing regression of electronic energies, we can measure the root mean square error associated with the density predictions carried out on any given test set as a fraction of the standard deviation of the reference test densities about  $\bar{\rho}_0(\mathbf{r})$, i.e., 
\begin{equation}\label{eq:error-def}
    \% \text{RMSE} = \sqrt{\frac{\sum_A\int d\mathbf{r}\left|\rho^\text{ML}_A(\mathbf{r})-\rho^\text{QM}_A(\mathbf{r})\right|^2}{\sum_A\int d\mathbf{r}\left|\rho^\text{QM}_A(\mathbf{r})-\bar{\rho}_0(\mathbf{r})\right|^2}} \times 100
\end{equation}
This choice carries the advantage of providing a measure of the learning capability of the method, rather than an absolute estimation of the accuracy of the predicted density. In practice, Eq.~\eqref{eq:error-def} can be efficiently computed as follows:
\begin{equation}\label{eq:error-calc}
    \% \text{RMSE} = \sqrt{\frac{\sum_A\left(\bc^\text{ML}_A-\bc^\text{QM}_A\right)^T\mathbf{S}_A\left(\bc^\text{ML}_A-\bc^\text{QM}_A\right)}{\sum_A\left(\bc^\text{QM}_A-\bar{\bc}_0\right)^T\mathbf{S}_A\left(\bc^\text{QM}_A-\bar{\bc}_0\right)}} \times 100
\end{equation}
where $\bar{\bc}_0$ is the vector of spherical ($\lambda=0$) coefficients associated with the expansion of the average density $\bar{\rho}_0(\mathbf{r})$. The overlap matrices $\bS_A$ of the test structures are set to the identity matrix when considering the prediction of orthogonal coefficients. 

To provide an estimate of the absolute accuracy of the predicted density we use the mean absolute error (MAE) of the density as a fraction of the total electronic charge:
\begin{equation}\label{eq:mae}
    \% \text{MAE} = \frac{1}{N_e} \int d \boldsymbol{r}\, \left|\rho_\text{ML}(\boldsymbol{r})-\rho_\text{QM}(\boldsymbol{r})\right| \times 100,
\end{equation}
which is computed using FHI-AIMS's dense logarithmic real-space grids centred on each atom, ensuring an accurate integration of the density close to the nuclei.\cite{AIMS}

\section{Results}

\subsection{Electron density of bulk liquid water}
To test the enhanced SALTED framework, we begin by considering a dataset consisting of 1000 configurations of 32-molecule supercells of bulk liquid water. This example presents enough complexity to require training the model on a large number of atomic density components, hence also requiring a large value of $M$, highlighting the importance of the methodological advances discussed here. In the process, we also compare the performance of the method with the alternative orthogonal approach outlined in Sec.~II-C. The reference density coefficients are computed at the DFT/LDA level by projecting the Kohn-Sham electron density on a set of auxiliary Slater-type atomic functions as implemented in FHI-AIMS~\cite{ren+12njp}. This results in a density field that is represented using up to $\lambda=8$ spherical harmonics. The corresponding DF strategy and the assessment of the DF error is discussed in detail in Ref.~\cite{lewis+21jctc}. 

Given the additional complexity of treating an extended condensed-phase system, we adjust the SALTED model according to the spatial resolution of the radial functions involved in the expansion of the density. For that, several $\lambda$-SOAP kernels are initially computed for various Gaussian widths $\sigma$ and  radial cutoffs $r_\text{c}$, in such a way that a ratio of $r_\text{c}/\sigma=10$ is always conserved. In particular, we consider six different cutoffs which span $r_\text{c}=(2.0,3.0,4.0,5.0,6.0)$~{\AA} and compute RHKS descriptors as in Eq.~\eqref{eq:psi-def} using $M=100$ sparse atomic environments for each basis function type. The selection of the sparse set is carried out via a farthest point sampling (FPS) algorithm which makes use of the SOAP metric to measure the distance between pairs of atomic environments~\cite{imba+18jcp}. For each of the aforementioned models, we then perform orthogonal-learning predictions which are cross-validated for each type of atomic functions $(an \lambda)$ on a two-fold 500/500 random partition of the dataset. Finally, the models that give minimal errors for each basis function type are selected to construct an optimal representation of the atomic structure to be used both in orthogonal and non-orthogonal approaches. Note that we will use all throughout the same sparse environments selections for each $(an \lambda)$.

\begin{figure}[t!]
    \centering
    \includegraphics[width=8cm]{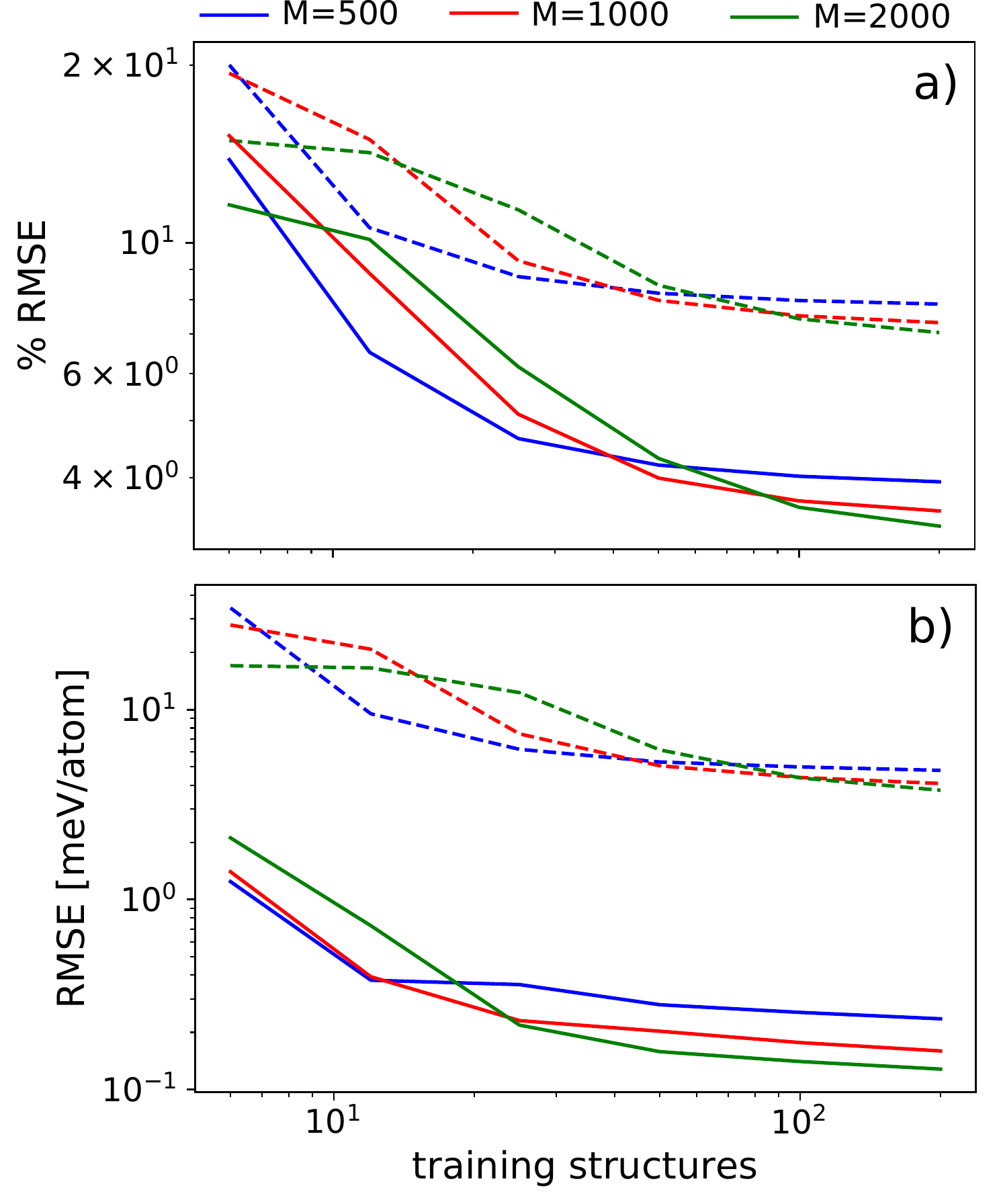}
    \caption{Learning curves associated with the prediction of 500 randomly selected liquid water structures using models constructed with an increasing number of sparse environments~$M$. (\textit{top}) \% RMSE of the predicted electron density as a function of the number of training structures. (\textit{bottom}) Absolute RMSE in meV/atom of the electronic energies obtained by feeding the predicted electron densities into the density functional used to generate the reference data. Full and dashed lines correspond to non-orthogonal and orthogonal learning models, respectively.}
    \label{fig:l-curves}
\end{figure}

Fig.~\ref{fig:l-curves}a reports learning curves for the electron-density prediction of 500 randomly selected liquid water frames, comparing regression exercises carried out at an increasing number of sparse atomic environments $M$.  Both for orthogonal and non-orthogonal learning approaches, we observe a saturation of the prediction accuracy for $M=2000$ sparse points and $N=200$ training structures, which carry a total of 19200 atomic density components. While the orthogonal approach yields optimal predictions associated with a RMSE of 7.0$\%$, making use of the fully coupled non-orthogonal framework allows us to bring the error down to $3.3\%$. Note that this level of accuracy corresponds to a mean absolute error of $~$0.36\% when measured as a fraction of the total electronic charge. In spite of the orthogonal approach being computationally more convenient to train, these results highlight the importance of adopting a fixed set of basis functions in order to solely focus regression effort on the density-based variations of the learning target. 

\subsection{Prediction of electronic-structure properties}

Having generated the reference data at the DFT level, the predicted densities can be used to compute the electronic energy of the system. Within the Kohn-Sham DFT formalism, however, one can only have direct access to the electrostatic and exchange-correlation energy, while the kinetic energy must be expressed in terms of the Kohn-Sham molecular orbitals. Therefore, we obtain the total energy by computing the Kohn-Sham potential as a functional of $\rho^\text{ML}(\mathbf{r})$ and solving the resulting eigenvalue problem. Doing so requires performing a single diagonalization step of the Kohn-Sham Hamiltonian, which can be done with almost linear-scaling cost thanks to dedicated libraries used in FHI-AIMS~\cite{ELSI2020}. As shown in Fig.~\ref{fig:l-curves}b, this procedure can in fact be used to yield extremely accurate electronic energies. In particular, for the maximum training set size adopted the predicted densities obtained through the non-orthogonal framework are associated with an error of 0.13meV/atom, greatly outperforming the accuracy that can be achieved through the orthogonal-learning framework. Interestingly, this level of accuracy is made possible by a large error cancellation between the kinetic energy $T$ and the potential energy $V$ of the system. As reported in the Supplementary Information (SI), computing separately the kinetic, electrostatic and exchange-correlation energy shows in fact that each of these individual terms are associated with errors that are from one to two orders of magnitude larger than those associated with the total electronic energy. The observed error cancellation is related to the fact that the first functional derivative of the total energy is zero when evaluated at the reference self-consistent density. In particular, when considering a functional expansion of the total energy $E$ in increasing powers of the density error $\Delta \rho$, the linear term in the error is expected to vanish at self-consistency, so that the leading order in the total energy error $\Delta E$ is in fact quadratic in $\Delta \rho$. A formal derivation together with a numerical test of the quadratic relation $\Delta E \propto \Delta \rho^2$ is reported in the SI.

\begin{figure}[t!]
    \centering
    \includegraphics[width=7cm]{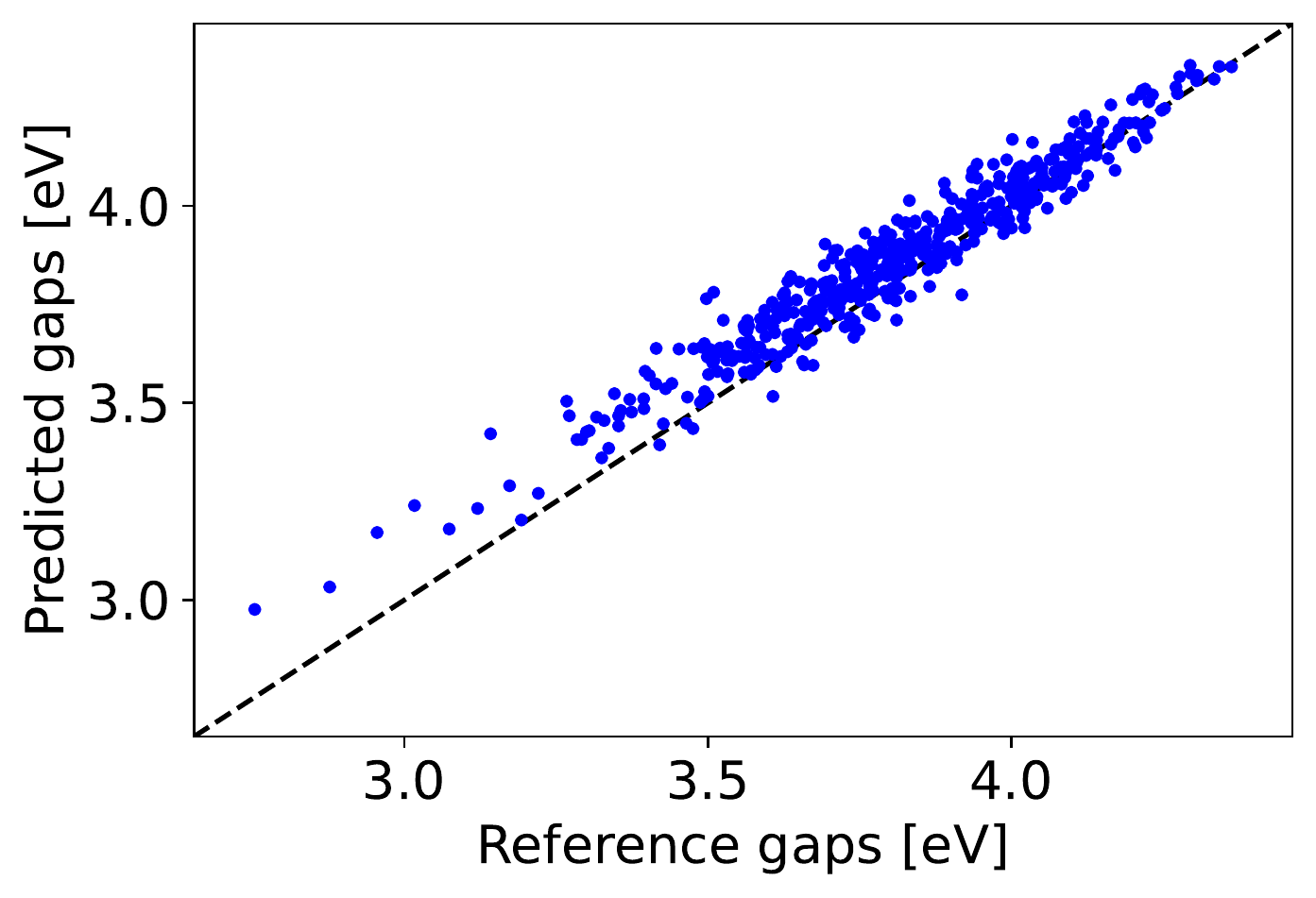}
    \caption{Correlation plot between the reference HOMO-LUMO gaps of 500 test water frames and those computed from the predicted electron density as obtained from a non-orthogonal learning exercise at $N=200$ and $M=2000$.}
    \label{fig:homo-lumo_gaps}
\end{figure}
In principle, being able to use the predicted density to solve the Khon-Sham eigenvalue problem allows us to access all kind of electronic-structure properties. In Fig.~\ref{fig:homo-lumo_gaps} we report an example of this by showing how the predicted HOMO-LUMO gaps of the 500 test water structures compare with the reference DFT values. Because of its non-local nature, the accurate prediction of this quantity is known to be a particularly hard problem~\cite{schu+19nc,nigam+21jcp}. As shown in the Figure, in spite of having learned a local property using local structural information only, our density-based predictions correlate well with the reference DFT values, resulting in a low root mean square error of 84meV. 
\begin{figure}[t!]
    \centering
    \includegraphics[width=8cm]{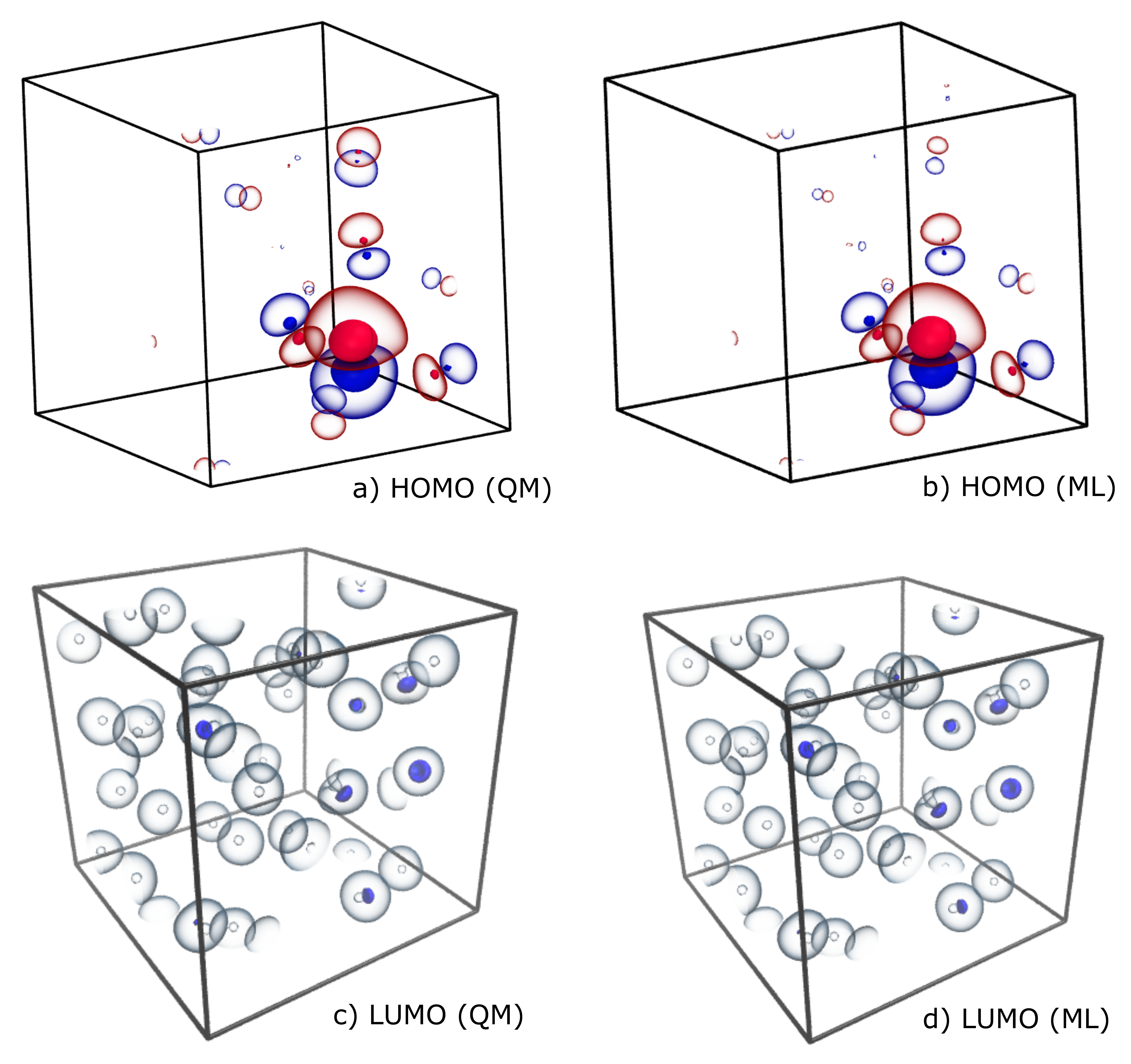}
   \caption{(\emph{top}) HOMO isosurfaces at $\pm$0.25 and $\pm$0.05 a.u. associated with the reference DFT calculation~(a) and those derived from the predicted density~(b). (\emph{bottom}) LUMO isosurfaces at 0.15 and 0.01 a.u. associated with the reference DFT calculation~(c) and those derived from the predicted density~(d).}
    \label{fig:homo-lumo}
\end{figure}
Furthermore, as shown in Fig.~\ref{fig:homo-lumo}, a 3D visualization of the HOMO-LUMO states of a given test water frame demonstrates an excellent agreement between the reference and predicted Kohn-Sham orbitals.

\subsection{The QM9 dataset}

As an even more challenging test, we learn the electron density of the QM9 dataset, containing more than 130,000 different molecules containing H, C, N, O and F atoms. Having demonstrated the superior performance of the non-orthogonal learning framework against the computationally more convenient L\"owdin approach, we will focus from here on exclusively on the former method. The vast chemical heterogeneity of the dataset requires dealing at the same time with a great number of sparse atomic environments and training structures, making it virtually impossible to use a naive implementation of SALTED where the regression exercise is carried out via the inversion of a kernel matrix. In testing the accuracy of our predictions we rely on the same partitioning of the dataset reported in Ref.~\cite{Jorgensen2021}, i.e., using the same random selection of 10,000 structures as a test set, while using the remaining configurations for training the model. In order to directly compare our results with what reported in Ref.~\cite{Jorgensen2021} we also make use of the same local environment definition by selecting a radial cutoff around the atoms of $r_\text{cut}=$4~{\AA}. For all learning exercises, we consider active set sizes that span $M=\{2500,5000,10000\}$. Note that having to deal with 5 different atomic types, the selection of sparse atomic environments plays a decisive role and it reflects the population of each atomic type in the training set. For instance, as very few structures contain fluoride atoms, only 29 fluoride environments are selected out of a total of $M=10000$.
\begin{figure}[t!]
    \centering
    \includegraphics[width=8cm]{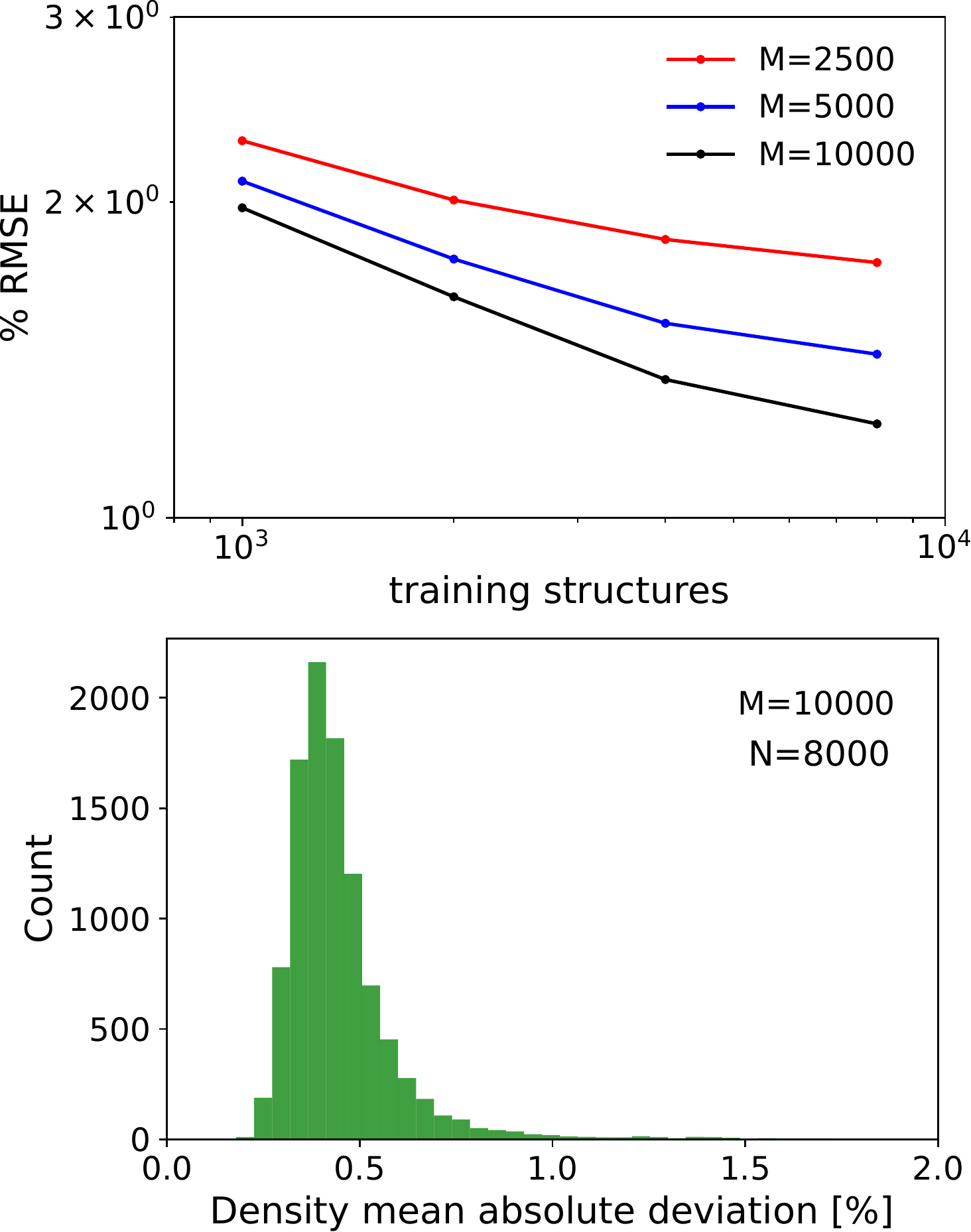}
   \caption{(\textit{top}) \% RMSE of the electron density prediction carried out on a total of 10,000 test molecules as a function of the number of training molecules at increasing values of $M$.
   (\textit{bottom}) Histogram of density absolute errors integrated in real space as a fraction of the total electronic charge computed at $M=10000$ and $N=8000$.}
    \label{fig:dens-qm9}
\end{figure}

\begin{figure}[t!]
    \centering
    \includegraphics[width=8cm]{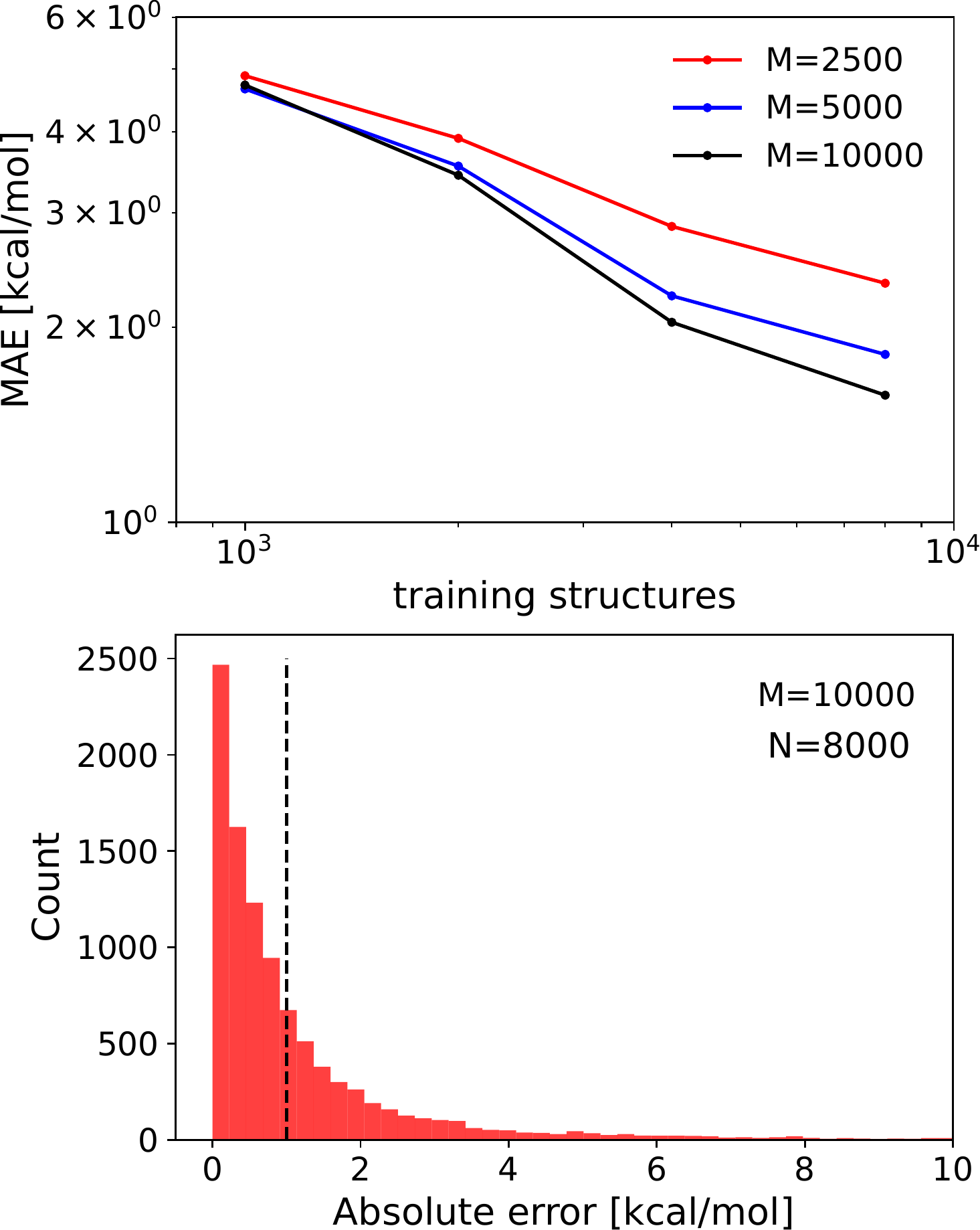}
   \caption{(\textit{top}) \% MAE of the total energy prediction as a function of the number of training molecules at a increasing values of $M$.
   (\textit{bottom}) Histogram of absolute energy deviations computed at $M=10000$ and $N=8000$. The vertical dashed line at 1 kcal/mol indicates the chemical accuracy threshold.}
    \label{fig:energ-qm9}
\end{figure}

Figure~\ref{fig:dens-qm9} reports learning curves for the prediction of the electron density using a maximum of 8000 randomly selected training structures reaching a small error of 1.22\% for $M=10000$. Although we observe superior learning performances when compared with the dataset of liquid water, the huge variance of the electron density in the QM9 dataset implies a lower absolute accuracy with respect to the previous example. In fact, as shown in the lower panel of the Figure, absolute density errors integrated on a real-space grid are distributed around a mean absolute error of 0.45\%, which is about 25\% larger than those observed for liquid water. When compared with the results of Ref.~\cite{Jorgensen2021} for the very same QM9 test molecules, we find that our errors are about 1.5 times larger; this is still remarkable when considering that we used only 6\% of the training set and that the reference calculations are performed for the all-electron density, rather than for a pseudo-valence density. In Figure~\ref{fig:energ-qm9} we also report learning curves and error histogram for the derived total energy predictions, showing a mean absolute error that is brought down to 1.57 kcal/mol, with 65\% of our predictions falling within chemical accuracy. Although not reported in the Figure for visual purposes, we find that there are 196 molecules with an error $>$10kcal/mol up to a maximum of $\sim$228kcal/mol. These outliers are likely to be associated with the lack of specific atomic environments in the training set, consistent with having used only a tiny fraction of the available data.

Learning curves for both densities and derived energies clearly show that there is still margin for improvement by increasing both $M$ and $N$, so that our method is likely to surpass the performance of Ref.~\cite{Jorgensen2021} when using the entire training set. However, for $N=8000$ and $M=10000$ we already find that performing the iterative minimization of the density loss function requires to keep in memory $\sim$10Tb of data, which are associated with the training overlaps and descriptors. This is no surprise, as the global problem dimensionality that accounts for all the basis functions of each of the 10000 sparse atomic environments is $D>2\times 10^6$ for this example, reflecting a large feature space of the RKHS descriptors. Going beyond these values of $N$ and $M$ and overcoming the need to use massive computational resources will be subject of future investigation.

\section{Conclusions}

The data-driven prediction of the electron density of a system is a challenging computational task which requires balancing the need for an optimal discretization of the scalar field with a regression model that is at the same time highly transferable and that can be easily integrated with electronic-structure programs. Adopting an atom-centered decomposition of the scalar field represents an optimal choice to simultaneously meet these requirements, but its integration within an efficient machine-learning algorithm carries exceptional computational challenges. We have shown how to tackle these challenges through an iterative minimization scheme of the density loss function which makes use of an equivariant sparse kernel approximation of the expansion coefficients. In doing so, we bypassed the need to invert regression matrices, increasing the capabilities of the method by 1-2 orders of magnitude with respect to previous implementations~\cite{lewis+21jctc}. This leap forward was enough to successfully tackle the regression of disordered condensed-phase systems such as liquid water and to achieve chemical accuracy on most of the test structures of the QM9 dataset with a handful of training structures. Further work should be undertaken to reduce the computational burden required to keep in memory all the overlap matrices and feature vectors needed during the training phase, thus allowing us to increase the training set size indefinitely. This could be done either by adopting better sampling algorithms that allow the selection of smaller active sets $M$ which are fine-tuned for each basis function type, or by implementing more efficient minimization strategies that  compute the gradient by loading the training data on-the-fly at each iteration.

An alternative approach, based on equivariant neural-networks that work on a similar density-fitting basis has recently been proposed~\cite{Rackers2022}. It directly uses the non-orthogonal expansion coefficients as the learning target. Although powerful, this strategy may be limited in accuracy, as it neglects the existing correlations between pairs of non-orthogonal coefficients~\cite{gris+19acscs}. 
In fact, a decoupling of the learning problem should rather follow an orthogonalization strategy analogous to that outlined in the present paper. We have shown that this strategy is outperformed by the (fully coupled) non-orthogonal framework, especially when using the predicted density to compute derived electronic energies.

As long as a single diagonalization step of the Kohn-Sham Hamiltonian can be performed at low computational cost~\cite{ELSI2020}, the possibility of predicting the electron density to access all kinda of ground-state electronic-structure properties is particularly attractive, as it sidesteps all intermediate learning exercises that might be carried out to predict quantities such as the density of states, band-gaps, wavefunctions and the Hamiltonian matrix itself. Moreover, the presented protocol could be used as an alternative to standard machine-learning potentials, where the total energy is learned and predicted directly from the atomic coordinates; while the computational cost of going through the density is necessarily higher, it allows one to more easily capture long-range effects that come from a more or less local integral of $\rho$ and, importantly, incorporates the permanent electrostatics exactly. As already shown in Ref.~\cite{lewis+21jctc}, this is extremely convenient when predicting the properties of systems that are much larger than those used during the training phase.

In this respect, a relevant next step will be investigating the role of long-range descriptors in complementing the capabilities of the method to predict all kind of physical interactions. For example, making use of far-field representations such as LODE~\cite{grisafi2021cs} to learn the electron density would help when reproducing long-range phenomena such as induced polarization and charge-transfer effects, which could not be captured through a local learning of the electron density around the atoms.
Finally, the accurate prediction of $\rho$ could be used to feed existing orbital-free kinetic-energy functionals, which are typically limited by the quality of the density, as well as to correct specific classes of KS-DFT calculations which present pathological density-driven errors~\cite{Sim2018,Sim2022}.

\section{Acknowledgments}
The authors thank Michael J. Willatt for an in-depth discussion of the orthogonal learning framework. We thank Peter B. J{\o}rgensen and Arghya Bhowmik for sharing the QM9 test configurations used as benchmark. AG acknowledges funding from the Swiss National Science Foundation. A.M.L. is supported by the Alexander von Humboldt Foundation. MC acknowledges funding from the European Research Council (ERC) under the European Union’s Horizon 2020 research and innovation programme (grant agreement No 101001890-FIAMMA).

\section{Source code and data availability}
An open-source implementation of the method, together with the dataset of liquid water, can be downloaded at \url{https://github.com/andreagrisafi/SALTED}.

\section{References}
\bibliographystyle{iopart-num}

\begin{thebibliography}{10}
\expandafter\ifx\csname url\endcsname\relax
  \def\url#1{{\tt #1}}\fi
\expandafter\ifx\csname urlprefix\endcsname\relax\def\urlprefix{URL }\fi
\providecommand{\eprint}[2][]{\url{#2}}

\bibitem{harmann+20nc}
Hermann J, Schätzle Z and Noé F 2020 {\em Nature Chemistry\/} {\bf 12}
  891--897 ISSN 1755-4349

\bibitem{Westermayr2021}
Westermayr J, Gastegger M, Sch\"utt K~T and Maurer R~J 2021 {\em The Journal of
  Chemical Physics\/} {\bf 154} 230903

\bibitem{kirkpatrick+21science}
Kirkpatrick J, McMorrow B, Turban D~H~P, Gaunt A~L, Spencer J~S, Matthews
  A~G~D~G, Obika A, Thiry L, Fortunato M, Pfau D, Castellanos L~R, Petersen S,
  Nelson A~W~R, Kohli P, Mori-Sánchez P, Hassabis D and Cohen A~J 2021 {\em
  Science\/} {\bf 374} 1385--1389

\bibitem{Kulik2022}
Kulik H, Hammerschmidt T, Schmidt J, Botti S, Marques M~A~L, Boley M, Scheffler
  M, Todorovi\'c M, Rinke P, Oses C, Smolyanyuk A, Curtarolo S, Tkatchenko A,
  Bartok A, Manzhos S, Ihara M, Carrington T, Behler J, Isayev O, Veit M,
  Grisafi A, Nigam J, Ceriotti M, Sch\"utt K~T, Westermayr J, Gastegger M,
  Maurer R, Kalita B, Burke K, Nagai R, Akashi R, Sugino O, Hermann J, No\'e F,
  Pilati S, Draxl C, Kuban M, Rigamonti S, Scheidgen M, Esters M, Hicks D,
  Toher C, Balachandran P, Tamblyn I, Whitelam S, Bellinger C and Ghiringhelli
  L~M 2022 {\em Electronic Structure\/}

\bibitem{bere+15jctc}
Bereau T, Andrienko D and Von~Lilienfeld O~A 2015 {\em J. Chem. Theory
  Comput.\/} {\bf 11} 3225--3233 ISSN 15499626

\bibitem{Staacke2022}
Staacke C~G, Wengert S, Kunkel C, Cs{\'{a}}nyi G, Reuter K and Margraf J~T 2022
  {\em Machine Learning: Science and Technology\/} {\bf 3} 015032

\bibitem{veit+20jcp}
Veit M, Wilkins D~M, Yang Y, DiStasio R~A and Ceriotti M 2020 {\em J. Chem.
  Phys.\/} {\bf 153} 024113 ISSN 0021-9606, 1089-7690

\bibitem{raim+19njp}
Raimbault N, Grisafi A, Ceriotti M and Rossi M 2019 {\em New J. Phys.\/} {\bf
  21} 105001 ISSN 1367-2630

\bibitem{Pilania2017}
Pilania G, Gubernatis J and Lookman T 2017 {\em Computational Materials
  Science\/} {\bf 129} 156--163 ISSN 0927-0256

\bibitem{mahmoud+20prb}
Ben~Mahmoud C, Anelli A, Cs\'anyi G and Ceriotti M 2020 {\em Phys. Rev. B\/}
  {\bf 102}(23) 235130

\bibitem{Westermayr2021cr}
Westermayr J and Marquetand P 2021 {\em Chemical Reviews\/} {\bf 121}
  9873--9926

\bibitem{snyd+12prl}
Snyder J~C, Rupp M, Hansen K, M{\"u}ller K~R and Burke K 2012 {\em Phys. Rev.
  Lett.\/} {\bf 108} 253002 ISSN 0031-9007

\bibitem{meye+20jctc}
Meyer R, Weichselbaum M and Hauser A~W 2020 {\em Journal of Chemical Theory and
  Computation\/} {\bf 16} 5685--5694

\bibitem{Ryczko2022}
Ryczko K, Wetzel S~J, Melko R~G and Tamblyn I 2022 {\em Journal of Chemical
  Theory and Computation\/} {\bf 18} 1122--1128

\bibitem{Zhou2019}
Zhou Y, Wu J, Chen S and Chen G 2019 {\em The Journal of Physical Chemistry
  Letters\/} {\bf 10} 7264--7269

\bibitem{Schmidt2019}
Schmidt J, Benavides-Riveros C~L and Marques M~A~L 2019 {\em The Journal of
  Physical Chemistry Letters\/} {\bf 10} 6425--6431

\bibitem{Suzuki2020}
Suzuki Y, Nagai R and Haruyama J 2020 {\em Phys. Rev. A\/} {\bf 101}(5) 050501

\bibitem{welb+18jctc}
Welborn M, Cheng L and Miller T~F 2018 {\em J. Chem. Theory Comput.\/} {\bf 14}
  4772--4779 ISSN 1549-9618

\bibitem{schu+19nc}
Sch{\"u}tt K~T, Gastegger M, Tkatchenko A, M{\"u}ller K~R and Maurer R~J 2019
  {\em Nat Commun\/} {\bf 10} 5024 ISSN 2041-1723

\bibitem{Gastegger2020}
Gastegger M, McSloy A, Luya M, Sch\"utt K~T and Maurer R~J 2020 {\em The
  Journal of Chemical Physics\/} {\bf 153} 044123

\bibitem{west-maur21cs}
Westermayr J and Maurer R~J 2021 {\em Chemical Science\/} {\bf 12} 10755--10764
  ISSN 2041-6520, 2041-6539

\bibitem{nigam+21jcp}
Nigam J, Willatt M~J and Ceriotti M 2022 {\em The Journal of Chemical
  Physics\/} {\bf 156} 014115

\bibitem{korits+01cr}
Koritsanszky T~S and Coppens P 2001 {\em Chemical Reviews\/} {\bf 101}
  1583--1628 pMID: 11709993

\bibitem{gris+19acscs}
Grisafi A, Fabrizio A, Meyer B, Wilkins D~M, Corminboeuf C and Ceriotti M 2019
  {\em ACS Cent. Sci.\/} {\bf 5} 57--64 ISSN 23747951

\bibitem{lewis+21jctc}
Lewis A~M, Grisafi A, Ceriotti M and Rossi M 2021 {\em Journal of Chemical
  Theory and Computation\/} {\bf 17} 7203--7214

\bibitem{alre+18cst}
Alred J~M, Bets K~V, Xie Y and Yakobson B~I 2018 {\em Composites Science and
  Technology\/} {\bf 166} 3--9 ISSN 02663538

\bibitem{chan+19npjcm}
Chandrasekaran A, Kamal D, Batra R, Kim C, Chen L and Ramprasad R 2019 {\em npj
  Comput Mater\/} {\bf 5} 22 ISSN 2057-3960

\bibitem{Jorgensen2020}
Jørgensen P~B and Bhowmik A 2020 Deepdft: Neural message passing network for
  accurate charge density prediction

\bibitem{Jorgensen2021}
Jørgensen P~B and Bhowmik A 2021 Graph neural networks for fast electron
  density estimation of molecules, liquids, and solids

\bibitem{broc+17nc}
Brockherde F, Vogt L, Li L, Tuckerman M~E, Burke K and M{\"u}ller K~R 2017 {\em
  Nat. Commun.\/} {\bf 8} 872 ISSN 20411723

\bibitem{bogo+20nc}
Bogojeski M, Vogt-Maranto L, Tuckerman M~E, Müller K~R and Burke K 2020 {\em
  Nat. Commun.\/} {\bf 11} 5223 ISSN 1

\bibitem{Rackers2022}
Rackers J~A, Tecot L, Geiger M and Smidt T~E 2022 Cracking the quantum scaling
  limit with machine learned electron densities

\bibitem{bearends1973}
Baerends E~J, Ellis D~E and Ros P 1973 {\em Chemical Physics\/} {\bf 2} 41--51
  ISSN 0301-0104

\bibitem{weigend+06pccp}
Weigend F 2006 {\em Phys. Chem. Chem. Phys.\/} {\bf 8}(9) 1057--1065

\bibitem{golze+17jctc}
Golze D, Iannuzzi M and Hutter J 2017 {\em Journal of Chemical Theory and
  Computation\/} {\bf 13} 2202--2214 pMID: 28383917

\bibitem{ren+12njp}
Ren X, Rinke P, Blum V, Wieferink J, Tkatchenko A, Sanfilippo A, Reuter K and
  Scheffler M 2012 {\em New J. Phys.\/} {\bf 14} 053020 ISSN 13672630

\bibitem{foster+09jmlr}
Foster L, Waagen A, Aijaz N, Hurley M, Luis A, Rinsky J, Satyavolu C, Way M~J,
  Gazis P and Srivastava A 2009 {\em Journal of Machine Learning Research\/}
  {\bf 10} 857--882

\bibitem{fabr+20chimia}
Fabrizio A, Briling K, Grisafi A and Corminboeuf C 2020 {\em CHIMIA
  International Journal for Chemistry\/} {\bf 74} 232--236 ISSN 0009-4293

\bibitem{fabr+19cs}
Fabrizio A, Grisafi A, Meyer B, Ceriotti M and Corminboeuf C 2019 {\em Chem.
  Sci.\/} {\bf 10} 9424 ISSN 2041-6520, 2041-6539

\bibitem{gris+18prl}
Grisafi A, Wilkins D~M, Cs{\'a}nyi G and Ceriotti M 2018 {\em Phys. Rev.
  Lett.\/} {\bf 120} 036002 ISSN 10797114

\bibitem{quin+05jmlr}
Qui\~{n}onero Candela J and Rasmussen C~E 2005 {\em J. Mach. Learn. Res.\/}
  {\bf 6} 1939–1959 ISSN 1532-4435

\bibitem{rasm05book}
Rasmussen C~E and Williams C~K~I 2005 {\em Gaussian {{Processes}} for {{Machine
  Learning}} ({{Adaptive Computation}} and {{Machine Learning}})\/} ({The MIT
  Press}) ISBN 0-262-18253-X

\bibitem{lowd1950jcp}
L{\"o}wdin P~O 1950 {\em J. Chem. Phys.\/} {\bf 18} 365--375

\bibitem{almof1991aqc}
Almlöf J and Taylor P~R 1991 Atomic natural orbital (ano) basis sets for
  quantum chemical calculations ({\em Advances in Quantum Chemistry\/} vol~22)
  ed Löwdin P~O, Sabin J~R and Zerner M~C (Academic Press) pp 301--373

\bibitem{marz-vand97prb}
Marzari N and Vanderbilt D 1997 {\em Phys. Rev. B\/} {\bf 56} 12847--12865 ISSN
  0163-1829

\bibitem{Marzari2003psik}
Marzari N, Souza I and Vanderbilt D 2003 {\em Highlight of the Month, Psi-K
  Newsletter\/} {\bf 57} 129

\bibitem{AIMS}
Blum V, Gehrke R, Hanke F, Havu P, Havu V, Ren X, Reuter K and Scheffler M 2009
  {\em Comput. Phys. Commun.\/} {\bf 180} 2175--2196 ISSN 00104655

\bibitem{imba+18jcp}
Imbalzano G, Anelli A, Giofr{\'e} D, Klees S, Behler J and Ceriotti M 2018 {\em
  J. Chem. Phys.\/} {\bf 148} 241730 ISSN 00219606

\bibitem{ELSI2020}
Yu V~W, Campos C, Dawson W, García A, Havu V, Hourahine B, Huhn W~P, Jacquelin
  M, Jia W, Keçeli M, Laasner R, Li Y, Lin L, Lu J, Moussa J, Roman J~E,
  Álvaro Vázquez-Mayagoitia, Yang C and Blum V 2020 {\em Computer Physics
  Communications\/} {\bf 256} 107459 ISSN 0010-4655

\bibitem{grisafi2021cs}
Grisafi A, Nigam J and Ceriotti M 2021 {\em Chem. Sci.\/} {\bf 12}(6)
  2078--2090

\bibitem{Sim2018}
Sim E, Song S and Burke K 2018 {\em The Journal of Physical Chemistry
  Letters\/} {\bf 9} 6385--6392

\bibitem{Sim2022}
Sim E, Song S, Vuckovic S and Burke K 2022 {\em Journal of the American
  Chemical Society\/} {\bf 144} 6625--6639

\end{thebibliography}
\providecommand{\newblock}{}

\end{document}